\documentclass[10pt]{article}

\usepackage{graphicx}
\usepackage{subcaption}

\usepackage[a4paper, left=35mm,right=35mm,top=34mm,bottom=34mm]{geometry}
\usepackage[utf8]{inputenc}
\usepackage[T1]{fontenc}
\usepackage[english]{babel}

\usepackage{enumerate}
\usepackage{graphicx}
\usepackage{hyperref}
\hypersetup{
    colorlinks=true,
    linkcolor=blue,
    filecolor=magenta,
    urlcolor=cyan,
}

\usepackage{mathtools,amsthm,amssymb,amsfonts}
\usepackage{algorithm}
\usepackage{algpseudocode}
\makeatletter
\def\thm@space@setup{
  \thm@preskip=10pt \thm@postskip=10pt
}
\makeatother
\theoremstyle{plain}

\theoremstyle{plain}

\theoremstyle{definition}

\theoremstyle{definition}

\theoremstyle{remark}

\theoremstyle{remark}

\usepackage{caption}
\usepackage{booktabs}
\usepackage{listings}
\usepackage{color}
\definecolor{dkgreen}{rgb}{0,0.6,0}
\definecolor{gray}{rgb}{0.5,0.5,0.5}
\definecolor{mauve}{rgb}{0.58,0,0.82}
\usepackage{enumitem}

\newcommand{\email}[1]{\protect\href{mailto:#1}{#1}}

\usepackage{xcolor}[2.11]
\colorlet{inlinkcolor}{green!50!black}
\colorlet{exlinkcolor}{red!50!black}
\if@hidelinks
\hypersetup{ hidelinks = true }
\else
\hypersetup{
  colorlinks = true,
  allcolors = inlinkcolor,
  urlcolor = exlinkcolor,
}
\fi

\newenvironment{@abssec}[1]{
        \vspace{.05in}\parindent .0in
        {\upshape\bfseries #1. }\ignorespaces
    }
    {\par\vspace{.1in}}
\renewenvironment{abstract}{\begin{@abssec}{\abstractname}}{\end{@abssec}}
\newenvironment{keywords}{\begin{@abssec}{Keywords}}{\end{@abssec}}

\usepackage{fancyhdr}

\author{
  {\normalsize Mengchen Wang }\thanks{CentraleSup\'elec, Universit\'e Paris-Saclay, 3 rue Joliot Curie, 91190 Gif-sur-Yvette, France
  (\email{frederic.magoules@hotmail.com})}  \thanks{VENISE Team, LIMSI-CNRS, Universit\'e Paris-Sud, Universit\'e Paris-Saclay, Orsay, France
  (\email{firstname.lastname@limsi.fr})}
  \and
  {\normalsize Nicolas F\'erey\footnotemark[2]}
  \and
  {\normalsize Patrick Bourdot\footnotemark[2]}
  \and
  {\normalsize Fr\'ed\'eric Magoul\`es\footnotemark[1]}
}
\title{Interactive 3D fluid simulation: steering the simulation in progress using Lattice Boltzmann  Method}
\date{}

\begin{document}
\maketitle
\thispagestyle{fancy}

\begin{abstract}
This paper describes a work in progress about software and hardware architecture to steer and control an ongoing fluid simulation in a context of a serious game application. We propose to use the Lattice Boltzmann Method as the simulation approach considering that it can provide fully parallel algorithms to reach interactive time and because it's easier to change parameters while the simulation is in progress remaining physically relevant than more classical simulation approaches. We  describe which parameters we can modify and how we solve technical issues of interactive steering and we finally show an application of our interactive fluid simulation approach of water dam phenomena.
\end{abstract}


\section{Introduction}
Steering a fluid simulation in progress requires to combine simulation, interaction and visualization tools. Usually these tools are sequentially used analysing simulation results at the end of computation using visualization tools without interaction features, except navigation around the fluid or only changing rendering properties or visualization modalities. In this paper we will discuss how we can interact and steer the fluid simulation in progress based on Lattice Boltzmann Method.

\section{INTERACTIVE FLUID SIMULATION WITH LBM}
In the last two decades, the Lattice Boltzmann Method has become very popular\cite{yeomans2006mesoscale} because it can provide fully parallel algorithms especially those adapted to many cores architecture such as GPU.

We choose the Lattice Boltzmann Method (LBM) for the simulation part of our system for the following reasons. On the one hand we need a high performance computation method to reach interactive time, i.e. providing several timestep per second to see the impact on interaction of user during a simulation in progress.  On the other hand, the final goal of our research is to develop a serious game targeting multi-phase fluid simulations and LBM is especially adapted in this context. Morover LBM are also less sensitive to parameters change during simulation allowing to steer the simulation during computation, as it is explained below.

The model of LBM used is the Bhatnagar–Gross–Krook collision model (BGK) \cite{he1997theory} with single-relaxation-time:
\begin{equation}
f_{i}(x+c_{i}\Delta t, t+\Delta t) - f_{i}(x,t)=-\frac{1}{\tau }(f_{i}(x,t)-f_{i}^{(0)}(x,t))
\end{equation}
In this equation $f_{i}^{(0)}$ is the equilibrium distribution function at x,t and $\tau$ is the time relaxation parameter\cite{perumal2008simulation}.
\subsection{Boundary conditions}
Boundary conditions and initial conditions are crucial in CFD simulations. To implement boundary conditions in LBM, we need to translate given information from macroscopic variables to particle distribution function\cite{noble1995consistent}. Since all the data are saved in particle distribution function, we can change boundary conditions at any time of the simulation. Technically, we can change all the boundary conditions by changing particle distribution function. For example, we can change the wall and modify the fluid in contact with the wall. We can also give a velocity of the wall and make the fluid in contact has the same velocity with the wall. This kind of solid objects are treated with stick boundary conditions, with the bounce-back on the links (BBL) \cite{ladd1994numerical}.
\subsection{Communication protocol}
Simulations and visualization/interaction are usually used separately. Both the simulation and interaction are time-consuming jobs for computers, reducing performance on both sides. This may cause latency that have to be avoid for qualitative interactive environment, and because latency causes cybersickness in the case of immersive context of serious game. \cite{pausch1992literature}.

To get a better performance, we chose to design a network API to implement a distributed architecture approach. One machine runs the simulation, and another machine runs the visualization and interaction. Since we have two machines and we want them to work interactively, we need them to efficiently communicate and exchange huge data with low latency. The simulation should be able to receive command from users at any time.

\begin{figure}[!t]
\centering
\includegraphics [width=4in]{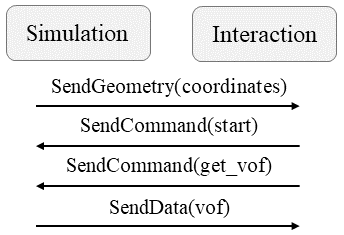}

\caption{Example of using communication API to launch a simulation and transmit volume of fraction(VOF)}
\label{fig19}
\end{figure}

This API is based on TCP protocol, and coded in C++. To make this API cross-platform, basic functions such as connection/send have been coded both for Windows and Linux as the TCP protocol has some differences in different platforms. These functions are completely coded in a encapsulating way. Then we design the API functions, there are no difference between platforms. We can compile this API to dynamic link library(DLL) for other languages like C\#. In this way connect two machines on different platforms and different languages. In our case, the simulation is coded in C++ under Linux, the interaction is realized with Unity under Windows. This API can also be used to connect other simulation libraries like OpenFOAM with other game engines. Fig.\ref{fig19} shows an example of using this API to transmit the simulation geometry, launch the simulation,  and transmit the volume fraction data. The commands should be defined in the simulation and interaction software. For sending data, data are sent in binary code. In this way we can send any type of data with passing the variable and the length of the data to the API.

Before every iteration or several iterations, simulation will check the received command. If the received command is empty, which means no command has been sent, the simulation goes on. If the simulation receives some command, it will treat the command, for example move a wall, changing only the cells concerned by the command and and keeping previous simulation step data for the others ones.

\subsection{Change boundary conditions during the simulation}
With LBM, it is possible to change the parameters as we need during the simulation. We can change the walls by simply setting the corresponding cells as fluid or wall. Then we only need to reinitialize the cells in contact with the new wall. To modify the inlet or outlet, we can simply change the parameters on the cells at the place of inlet/outlet. This makes it possible to only transfer very little data between two machines. Only modified cells data in the 3D scene will be sent to the simulation, and user have a visual feedback of the edited boundary conditions or cells content inside the 3D environment.

\begin{figure}[!t]
\centering
\includegraphics [width=4in]{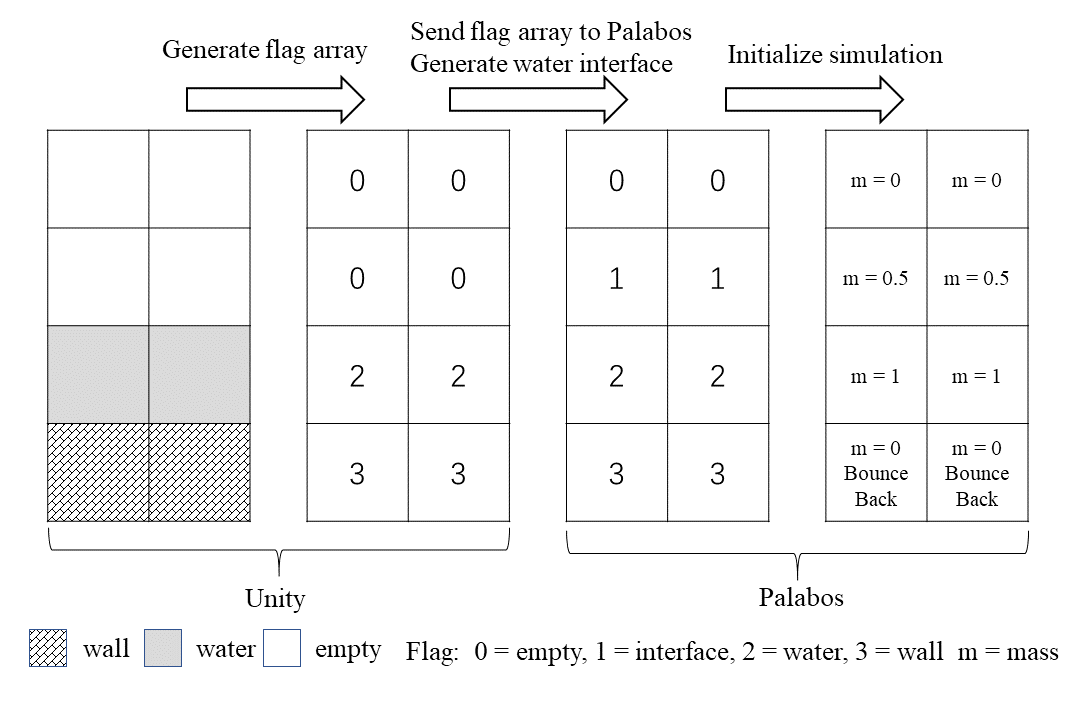}

\caption{Process of setting initial conditions}
\label{fig8}
\end{figure}

Fig.\ref{fig8} shows a proposition of how to deal with the initial condition. In Unity, the property of each cell is showed to the user with colors or textures. These data can be saved in an array flags which stands for wall, water, etc. Then the array will be sent to the simulation and initialize the initial conditions with the array of flags.

\section{3D ENVIRONMENT VISUALIZATION}
In case of 3D fluid data, there are a large number of different techniques for flow visualization. The choice depends on the circumstance:  the purpose of the visualization, the result of the simulations or experiments and the objectives of the analysis.

Flow visualization is a subfield of scientific visualization especially dedicated to render flow. It has a wide application, like weather forecast, aerodynamics, airplay design, etc. This means it is widely used in different aspects. So, the users may have different purpose and there are many flow visualization resolutions.

Flow data is normally a set of positions, velocity vector, pression... There is a lot of flow visualization techniques aiming to present the whole flow data or partially in a comprehensible way, described for example in the classification of Post, Frits H., et al. \cite{post2002feature}.
\subsection{Direct flow visualization (Direct FlowVis)}
 One chosen approach is direct flow visualization. This technique aims to present the data in an explicit way as the data are.  There are also many kinds of Direct FlowVis, such as Direct FlowVis 2D or 3D including Color coding, Arrow plots and Hybrid direct FlowVis. Direct FlowVis on slices or boundaries including Color coding and Arrow plots. An example of Direct flow visualization is using this technique to visualize the particles of micro fluid\cite{roberts2007direct}. We used these techniques to visualize pressure, velocity and interface like water surface. This category maybe the most important in our case as it shows to users in a direct way how the fluid evolve.
\subsection{Geometric Flow Visualization}
Geometric Flow Visualization is also an approach of flow visualization. This technique uses some geometric objects to present the potential information in the data. The shape of the objects is related to the information. Geometric Flow includes Contouring, Isosurfaces, Streamlets, Streamlines, Streaklines, timelines, and pathlines. We can use for example streamlines to trace the fluid movement. These techniques offer more information than direct flow visualization techniques.
\subsection{Texture-based visualization}
Texture-based visualization is another approach of the flow visualization. This technique provides dense spatial resolution images to the users. This technique presents the data to the users by rendering the data with some texture. Texture-based FlowVis including Spot noise, Line integral convolution, Oriented Line Integral Convolution. Cross advection and error diffusion \cite{botchen2005texture} are two examples of texture-based techniques to visualize uncertainty in time-dependent flow.

\subsection{Feature extraction}
Apart these standard flow visualization techniques, a very specific flow rendering. For example, when it concerns the vortex structures of a flow, such structure is not directly accessible in the result of a CFD simulation, while vortex has a great importance for both theoretical and practical research.  \cite{jiang2002novel} proposed a simple and efficient vortex core region detection algorithm based on ideas derived from combinatorial topology. We can for example extract the centers of vortex in the fluid, or other specific features that the user needs.

Amount all these visualization techniques, we must choose some more accessible techniques. In a interactive scenario, users do not have much time to analyze data, so the showed results must be very easy to understand.

\begin{figure}
\centering
\includegraphics [width=4in]{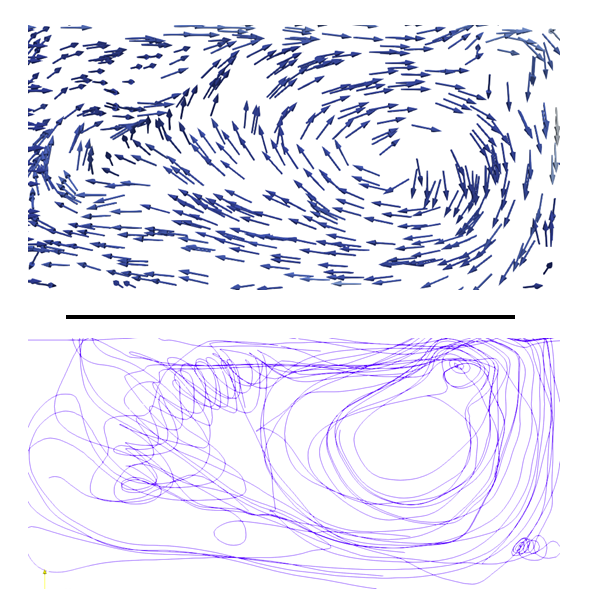}\\
\caption{Comparison between direct(up) and geometric(down) flow visualization}
\label{fig18}
\end{figure}

\section{INTERACTION IN 3D ENVIRONMENT}
In this section I will focus on how to interact with objects in 3D scene.  The most common tasks are object selection and manipulation. The simplest way to select an object in a 3D environment using 3D mouse input is to use ray casting technique. This technique casts a ray perpendicular to the pixel on the screen and compute the intersection between this ray and the nearest object considered as the selected 3D object.

In fluid simulations, except selecting the solid object in the environment, fluid is the 3D field discretely stored on cells or points grid. If we want to choose a part of fluid in a 3D environment, we have to use efficient picking paradigm to select cells or points that have to be edited.

Manipulating objects is another task we can perform in a 3D scene, for example moving a whole wall in the CFD simulation instead of set it cell by cell. One way to achieve this is to use manipulators\cite{strauss2002design} to move the whole along x, y, z axis. Manipulators are defined as visible graphic representations displayed on the objects such as arrows, to allow user to choose the direction of move of a wall object composed of several cells or point.

Pointing and picking can also be improved by advanced 3D pointing techniques. Elmqvist and Fekete\cite{elmqvist2008semantic} developed an adaptation called semantic 3D pointing. This adaptation makes pointing more accurate by shrinking empty space and expanding target sizes in motor space. This technique may be very useful for our case, as the fluid field mesh is very fine. If we want to select a certain part of the simulation field, this technique increase the precision.

\section{AN APPROACH FOR INTERACTIVE FLUID SIMULATION}

\subsection{Hardware and software architecture}
For our interactive fluid simulation approach, we designed and implement a distributed hardware and software architecture. One machine is dedicated to simulation, and the other one runs a 3d scene including rendering and interaction features based on Unity 3D. On the contrary to work of Florian De Vuyst et al. \cite{devuyst2013} in which simulation and rendering process was done on the same GPU, we choose to design a generic network API to enable data exchange between simulation and rendering software component, in order to easily extends any existing simulation and visualisation tools with interactive simulation features.

Using this approach, users can interact with a simulation in progress in interactive time, modify the simulation conditions and parameters without stopping the simulation, or restart the simulation from the beginning with new parameters or boundary condition.

We choose \emph{Palabos} (http://www.palabos.org/) as simulation library, which is a free, open source software based on LBM. For the visualization and interaction, we choose to use Unity 3D. Unlike other tools as \emph{Paraview}, Unity makes user to quickly and easily develop custom rendering and interaction techniques needed to steer a simulation in progress.

\subsection{Proof of concept with a water dam}
\begin{figure}[!t]
\centering
\includegraphics [width=4in]{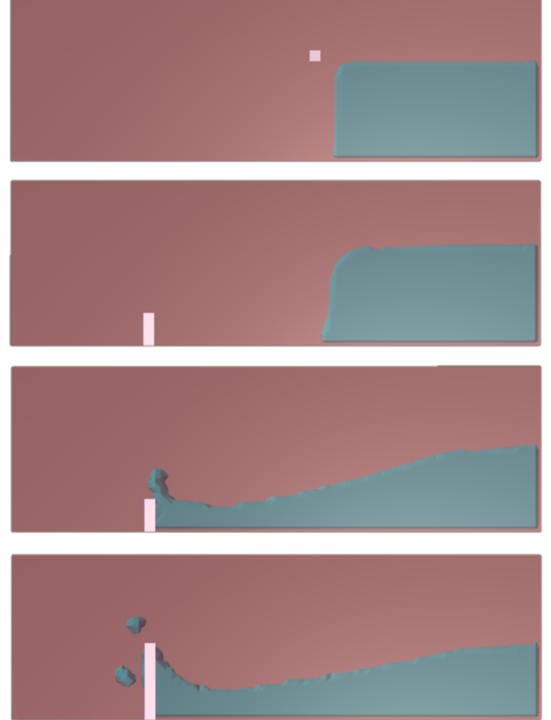}
\caption{Design of the dam during the simulation in progress}
\label{fig9}
\end{figure}

We made a benchmark to test the performance with the simulation of a 3D water dam. We only render and interact with a 2D cutting plane of 3D simulation inside \emph{Unity 3D}, and user can set and move the shape of the 3D dam as shown in the Fig.\ref{fig9}.

This simulation of the water goes cross the dam includes 206080 cells, and for every second a simple laptop can run around 20 iterations.

Reaching interactive time performance inside Unity 3D is important to avoid cybersickness \cite{laviola2000discussion}, and the mean refresh rate obtained with direct flow rendering of the 70K triangles mesh is about 60Hz, using a thread especially dedicated to network communication features.

\begin{table}[h]
\centering
\caption{Simulation with \emph{Palabos}}
\begin{tabular}{|c|c|}% 通过添加 | 来表示是否需要绘制竖线
\hline  % 在表格最上方绘制横线
Simulation field&$96\times30\times30$ (86400) cells\\
\hline
CPU for simulation&i5-4200M 3.1GHz\\
\hline
Calculation speed&~10 iterations per second\\
\hline
Result data type&Double(water interface mesh)\\
\hline
Result output rate&1Hz\\
\hline
Data transmitted&$\sim$200K doubles (800KB)\\
\hline
\end{tabular}
\end{table}

\begin{table}[h]
\centering
\caption{3D scene in Unity}
\begin{tabular}{|c|c|}% 通过添加 | 来表示是否需要绘制竖线
\hline
CPU for interaction&I7-8750u 4 cores 4.8GHz\\
\hline
GPU for interaction&GTX1050 max-Q\\
\hline
Unity scene refresh rate&60Hz\\
\hline
\end{tabular}
\end{table}
\section{Conclusion}
We designed a system which allows users to interact with a fluid simulation in progress targeting an application of serious game for pedagogical purposes. Our approach allows user to observe results of the simulation in progress in interactive time, and to directly observe the impact of modifying boundary condition or fluid parameters without restart the simulation. We designed an API for the network communication between the simulation and the interaction, allowing to be suitable to any simulation or visualisation tools.

We discussed about visualization and interaction techniques to manipulate and edit objects in  the 3D scene. The next step for us is to design and implement scenario in the context of serious game, addressing more complex phenomena with the goal to optimize some global parameters of the fluid, or to reach an given objective to the player, by modifying in interactive time the parameter and conditions of the simulations.

\bibliography{paper}
\bibliographystyle{abbrv}

\end{document}